\documentclass[preprint, a4paper, times, 11pt]{elsarticle}
\pdfoutput=1
\usepackage{amsmath}
\usepackage{ecrc}
\usepackage{listings}
\usepackage{graphicx}
\usepackage{epstopdf}
\graphicspath{{./images/}}
\DeclareGraphicsExtensions{{.eps}}

\volume{00}

\firstpage{1}

\journalname{NICTA}

\runauth{Shvartzshnaider, Ott}
\runtitle{Design For Change: Information-Centric Architecture to Support Agile Disaster Response}

\jid{procs}

\jnltitlelogo{NICTA}

\usepackage{amssymb}

\begin{document}

\begin{frontmatter}

\dochead{}

\title{Design For Change: Information-Centric Architecture to Support Agile Disaster Response}

\author[label1,label2]{Yan Shvartzshnaider}
\ead{Yan.Shvartzshnaider@sydney.edu.au}
\author[label2]{Maximilian Ott}
\ead{max.ott@nicta.com.au}

\address[label1]{University of Sydney, NSW, Australia}
\address[label2]{NICTA, Australian Technology Park, Eveleigh, NSW, Australia}

\begin{abstract}

This paper presents a case for the adoption of an information-centric architecture for a global disaster management system.  Drawing from a case study of the 2010/2011 Queensland floods,  we describe the challenges in providing every participant with relevant and actionable information. We use various examples to argue for a more flexible information dissemination framework which is designed from the ground up to minimise the effort needed to fix the unexpected and unavoidable information acquisition, quality, and dissemination challenges posed by any real disaster.

\end{abstract}

\end{frontmatter}

\newpage
\newlength{\picturewidth}
\setlength{\picturewidth}{0.75\columnwidth}
\section{Introduction}

Getting the right information to the right people is a crucial component in any  effective response to a disaster. Many information systems have been designed and deployed to provide, what is often referred to as a ``Common Relevant Operational Picture'' (CROP). CROP provides each participant with a small subset from a common information base  to allow efficient operation and coordination of their activities with minimum distraction and information overload.

However, no matter how well these systems have  been researched and planned, every disaster brings new surprises and quickly reveals the omission of important sources of information or breaks in the required information flows. We therefore argue that one of the most important characteristics of an information system is the ease of adding new sources of information, new ways of combining information to create new insights, and new ways of determining what is relevant to whom. 

Recent disasters saw the emergence of mash-ups as a quick method to blend various information sources into individual services. Unfortunately, in this case the information integration only happens at the presentation layer. While this is useful for a human, it does not lend itself very well to be folded back into a common information space.

Another common problem is the varying quality of information -- or lack of it -- on which decisions are based and new information is created. While this is unavoidable, maintaining the dependencies between information entities would allow us to more efficiently understand the impact of new or retracted information in the overall information base.

To fundamentally tackle this problem, we argue that we need to adopt an information-centric model instead of the current workflow-oriented one. Rather than defining and concentrating on who should be providing what to whom on a case-by-case bases, we need to develop a model based on semi-independent entities which consume information from a common information space  and contribute their actions and findings to the same. 

\begin{figure}
\centering
	\includegraphics[width=0.75\textwidth]{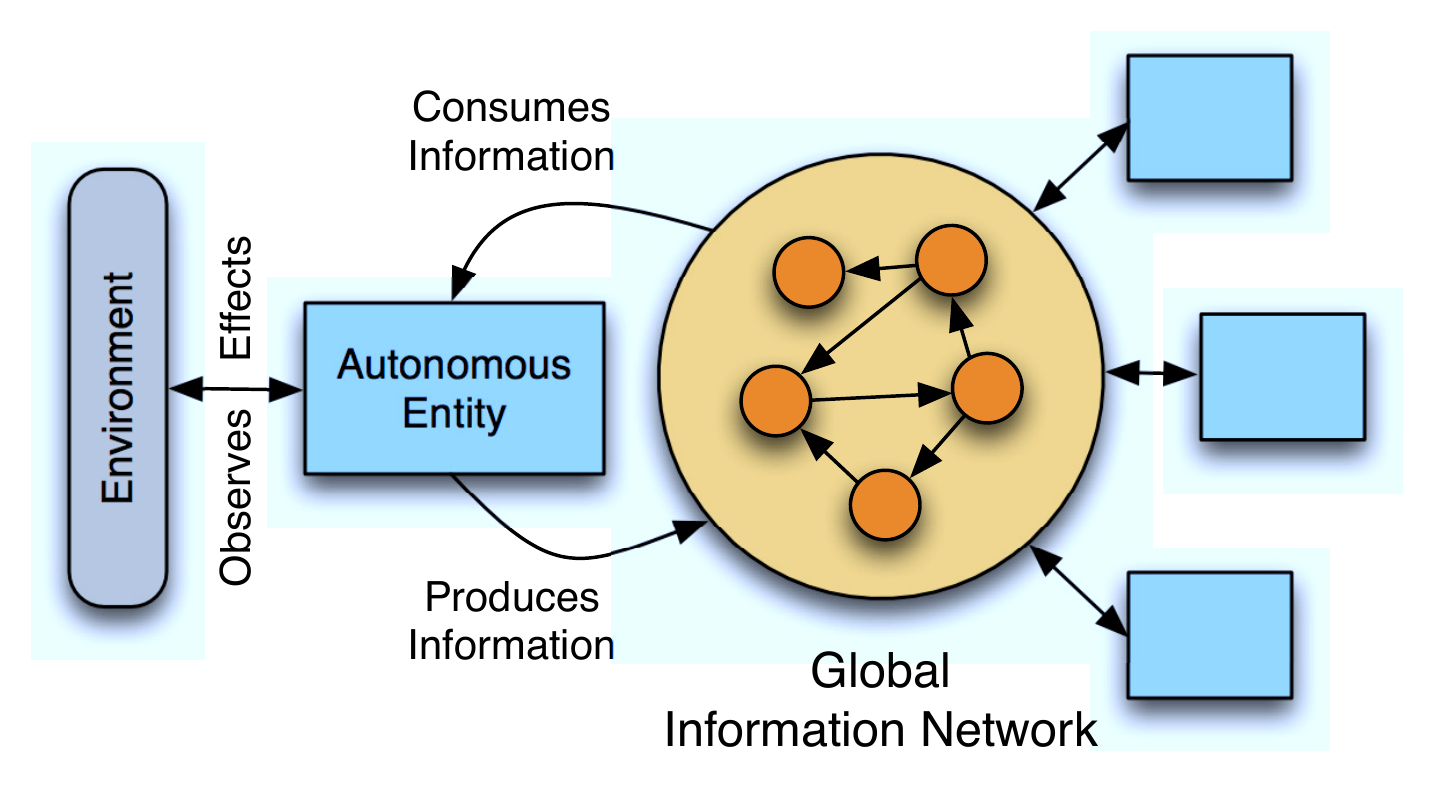}
\caption{Information Centric System Model.}
\label{model}
\end{figure}

We propose a Global Information Network (GIN)~\cite{GIN2011} that loosely couples autonomous entities as shown in Figure~\ref{model}, and quickly reacts to  new information provided by anyone. This model brings to the forefront the fact that a common information pool,  especially in a disaster scenario, will contain conflicting and simply wrong information. We believe that we need to cater for these cases from the beginning to achieve truly robust systems. 

While the Semantic Web with its formal foundation has not caught on with the broader developer community, the more pragmatic approach of the various ``Linked Data'' initiatives has resulted in a much quicker uptake \cite{lod_cloud}. Both approaches are based on graph-like representations.%

GIN is therefore based on a hypergraph abstraction that acts as a storage and information dissemination network for distributed, autonomous agents. GIN supports the basic operations of \textit{add} and \textit{map}. Contrary to traditional pub/sub system, a publication in GIN is essentially an addition to the persistent global graph, while a subscription is a mapping of the global graph to an application specific local sub-graph. The service only supports an ``add-only'' modus, which ensures that the full history of information discovery is kept, while the hypergraph supports annotation and provenance. The GIN itself is schema-agnostic and considers the semantic interpretation of labels of the vertices and edges as an end-to-end property. Our current GIN implementation is tolerant to network disruptions and will automatically propagate information when networks merge again. 

The rest of this paper is organised as follows. In the next section we present a case study into the recent Queensland floods from the perspective of information exchange between the involved entities. In Section \ref{icm}, we discuss an alternative information-centric model and list its key design goals. Section \ref{related_work} lists some highlights from the existing literature. Section \ref{GIN} briefly describes the Global Information Network and gives an example how it can be utilised by a disaster management system. Section \ref{challenge_and_progress} describes our current progress. Finally, we describe our future plans and conclude the paper.

\section{Disaster Management}

At any stage of  disaster management -- planning, response, or recovery -- information is of essence \cite{national1999reducing}. There is a constant information flow between various emergency agencies, the government and the public. This information is the base for critical decisions, issuance of warnings, forecasting of conditions, or logistics planning, just to name a few. 

To better understand the complex information flows and dependencies we briefly discuss the recent floods in Queensland, Australia.

\subsection{Case study: 2010/2011 Queensland Floods} \label{case_study}

The Australian State of Queensland is prone to flooding during the yearly wet season. The disaster management groups at the state, district, and local level follow a common disaster management plan for each phase of a disaster: prevention, preparedness, response and recovery~\cite{qdm_website}. Queensland's authorities are well equipped and have a lot of experience in handling disaster situations,  floods in particular.  Tropical Cyclone Tasha in December 2010 and a peak in the La Ni\~{n}a weather pattern caused prolonged and heavy rainfall over Queensland river catchments. At some stage, more than half of Queensland, an area more than twice as large as Japan, was affected by flooding. The floods killed approximately 40 people and caused roughly A\$\,1 billion in damages. The sheer scale of this revealed many breaks in the current Queensland Disaster Management framework leading to a State inquiry.  The following presents our subjective interpretation of the Queensland Flood Commission of Inquiry's interim report \cite{inquiry_report}.

\begin{figure*}[t]
\centering
	\includegraphics[trim =18cm 0cm 0cm 0mm, clip=true,width=1.5\textwidth]{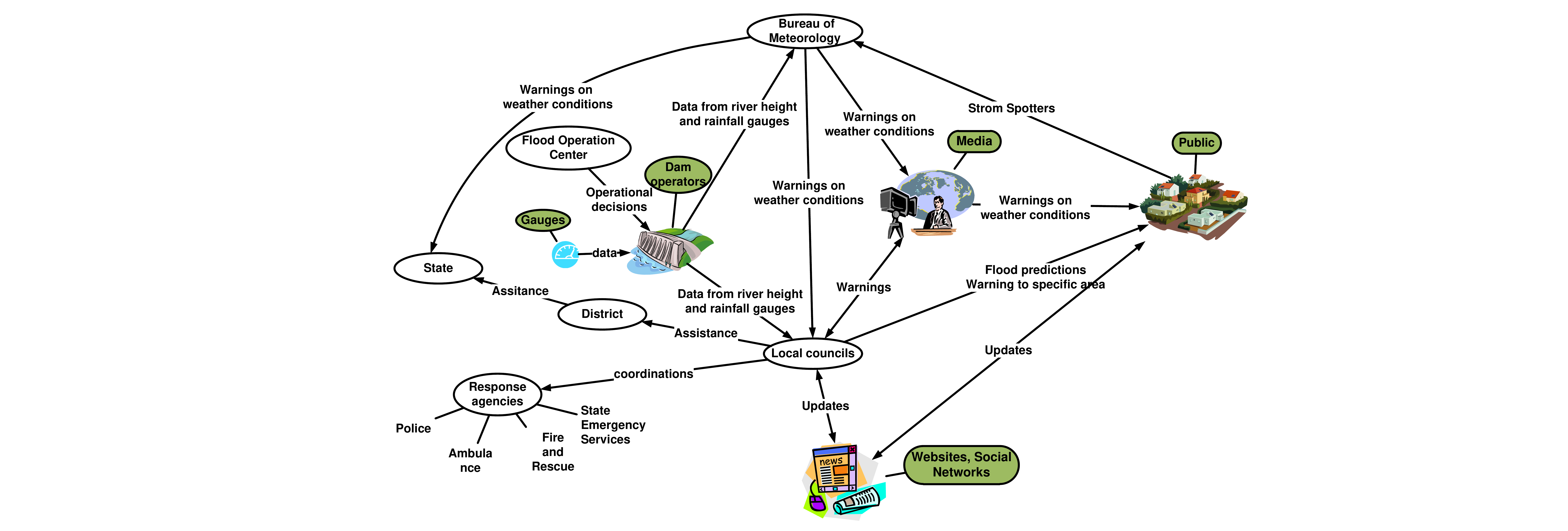}
\caption{Queensland Disaster Management Information Flow}
\label{info_flow}
\end{figure*}

As depicted in Figure \ref{info_flow} there are many agencies involved in the management of a disaster that require different information at varying granularity to make decisions and assess the current situation. Many of the agencies maintain their own, internal information system and contact external parties for additional or missing data. Even more so, in addition to information, human factors play a very crucial role in the current disaster management framework.

For example, dams play a crucial role in managing catchment runoff. In case of flood danger, the dam operators are responsible for setting up Flood Operation Centres (FOC) staffed by certified flood engineers tasked with managing dam operations. The engineers utilise the real-time flood monitoring system for flood forecasting. It generates hydrographs of runoff in a catchment based on the data from rainfall and water level gauges in the area.
In addition, the FOC tries to maintain a constant phone connection with the Bureau of Meteorology (BoM) forecasters. Based on all the information collected, the flood engineers decide on a specific gate operation strategy and produce ``situation reports''. Such reports contain information on recorded rainfall, lake level as well as the current and predicted release rates of the dams and their likely impact on the lakes and rivers down streams. The reports are sent via email to various agencies and the councils. The dam operator manager also edits the reports to produce a ``technical situation report'', which are sent to the Queensland Police, as well as the Department of Premier and Cabinet. The creation of the two reports can lead to confusion and wrong interpretation. 

Also of interest is the finding by the commission that the effectiveness of the information exchange between entities, such as the BOM forecaster and the flood engineer, very much depended on the closeness of their personal relationships. Shift changes which may not be visible to others, may substantially affect the quality of the information flow.

The local councils have the primary responsibility of managing the disaster on the ground.  They have to interpret the  gauge data from the dam operators,  and the  forecast and warnings from the BoM to determine the likelihood and extent of inundation of individual properties. Decisions to evacuate areas need to be communicated to the affected population in a timely manner. Not only do they need to contain enough information  to convey the danger and potential consequences, but they also need to consider other factors, such as non-English speakers or deaf communities. Considering the cognitive effort needed and the lack of support from underlying system, it is really not surprising that "some councils experienced delays when sending SMS alerts to residents, caused by the time taken to draft the text of the alert and identifying which residents should receive it" \cite{inquiry_report}. 

An independent review \cite{brisbane_report} of the Brisbane City Councils response to the floods determined that Facebook  and Twitter were used extensively to access information about the floods. For instance, the Australian Broadcast Corporation (ABC) quickly launched a Flood Crisis Map\footnote{http://queenslandfloods.crowdmap.com/} to ``crowd source'' information about road closures and availability of electricity. It also allowed people to indicate where help may be needed as many call centres experienced multi-hour queues. The Flood Crisis Map also experimented with trust levels on information. 

In summary, the disaster management framework, from an information dissemination point of view, primarily prescribes information flows between selected entities and heavily relies on humans to combine, interpret, and process the locally available information.

\section{Information-centric model}\label{icm}

In a disaster,  everyone is hungry for information.  In most disaster management systems the ``information is often produced from disparate sources and transmitted in whatever format the provider prefers, requiring significant effort to compile it into a form that provides a coherent picture or even thwarting integration altogether" \cite{national1999reducing}. 

The Web showed us the power of information integration; the recent disasters saw the emergence of web mash-ups as a quick way to aggregate related data from various sources. This type of flexible information integration is done at the presentation layer and is not represented in any way at the underlying system. 

We argue, the information-management system should  be flexible and dynamic enough to adapt to the evolving situation, facilitate quick integration of new sources of information and support the overall decision making. Next we present the design goals we believe such an  information-centric network architecture must address to provide suitable support for running on top future applications such as disaster management system.

\begin{figure*}[t]
\centering
	\includegraphics[trim =0cm 15cm 15cm 1mm, clip=true, width=1\textwidth]{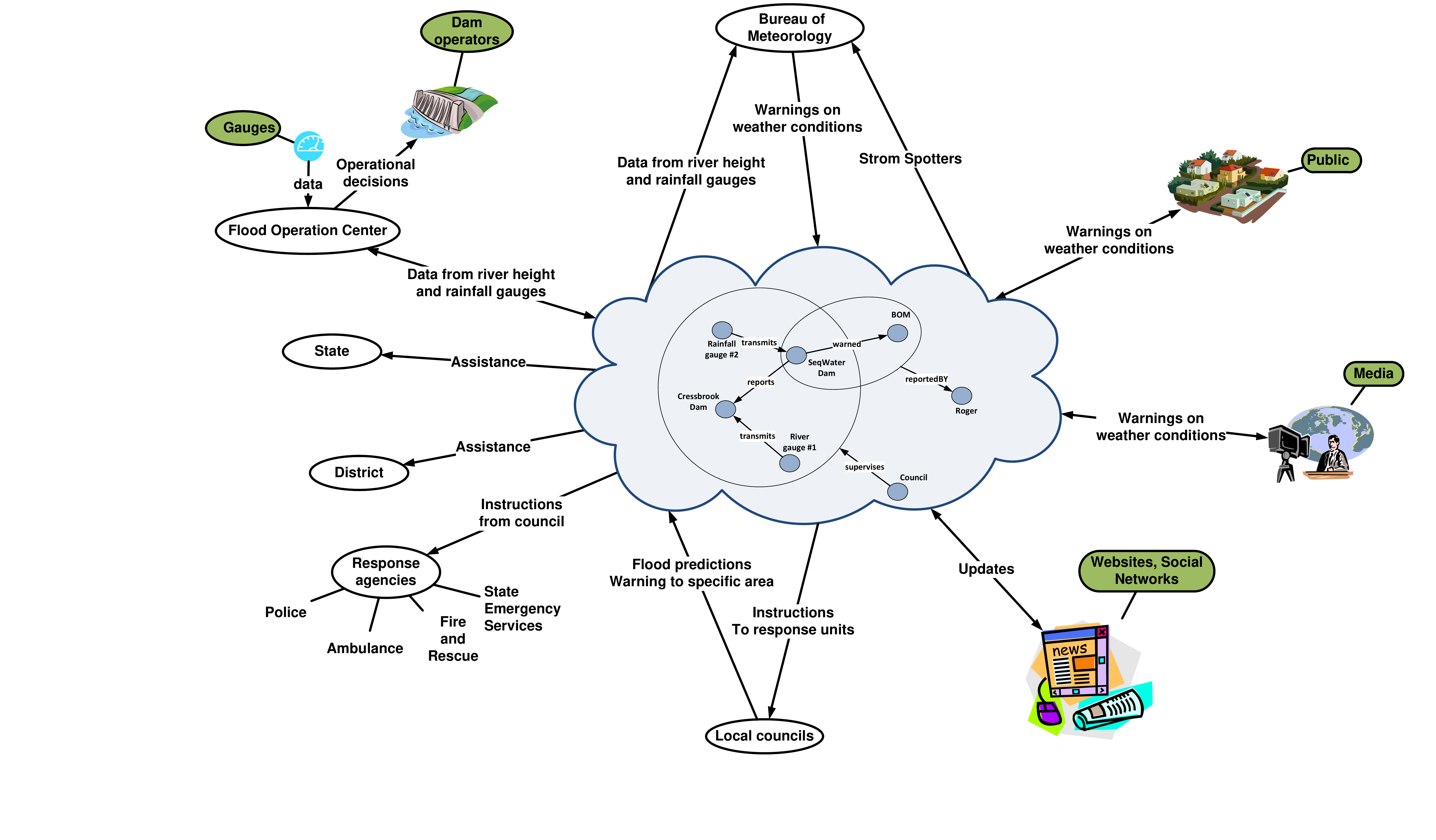}
\caption{GIN-based Information model}
\label{gin_based_model}
\end{figure*}

\subsection{Architectural Design goals}\label{design_goals}

In this section we present the design goals for an information-centric system. 
\medskip\\\textbf{Shared information space.} The current disaster management frameworks do not fully take advantage of the distributed information repositories in the different agencies \cite{national1999reducing}. As depicted by Figure \ref{gin_based_model} we propose establishing a globally shared space to which all the information will be added. This will allow all the participating parties to  have an access to the information. 
\medskip\\\textbf{Standardized information format.} The current frameworks often do not put a strong emphasis on the data formats. However, the justifications often confuse data format versus data model. We agree, that standardising data models is hard, agreeing on a standard data format -- or a small number thereof -- should be much easier. This will reduce development time as one can leverage many of the existing tools and libraries. It also reduces the development of ``adaptors'' between different model domains to concentrate on the model aspect. Standardised ``low level'' models such as RDF\footnote{Resource Description Framework} also provide support for systematic alignment of different information domains.  
\medskip\\\textbf{Asynchronous communication.} Disaster management often requires  many-to-many communication where the producer often does not know who would be interested in its output. The usual host-to-host communication paradigm is not a well suited to the requirements of the disaster situation. Multiple sources as well as concurrent and frequent queries for information call for a more scalable solution such as publish/subscribe.
\medskip\\\textbf{Transparently adding new providers.} It is important to have the ability to easily add  new sources. In addition to traditional sources of information, social networking services, such as Facebook, Twitter and others,  are becoming an important source of valid information. It is also extremely easy for ``community programmers'' to quickly develop and deploy services and applications to allow others to provide useful information in ``actionable'' formats. What is lacking is a mechanism to feed this into the global information pool.
\medskip\\\textbf{Context.} In a disaster situation the lack of context leads to confusion and potentially wrong decisions. Warnings issued to the public by agencies have little impact if they are not accompanied by ``more detailed information 
about flood locations and predictions, the location of evacuation centres and evacuation routes" \cite{inquiry_report}.  We argue that an information system needs to maintain the relationships among information entities leading to a graph abstraction which is able to capture these relationships and better reflect the context in which the information originated.
\medskip\\\textbf{Add-only.} Historical (long and short) records are very useful to calculate trends or spot inconsistencies. In fact, maintaining global consistency in a large and distributed and highly dynamic system is highly unrealistic. In contrast, an ``add only with timestamps'' mechanism will allow for temporal constancy within a sub graph. We should point out that due to errors in clock synchronisation,  timestamps only provide limited support for ordering. However, current time synchronisation protocols, coupled with the high precision clock underlying the readily available GPS system, time drift across servers can be assumed to be bound rather tightly.
\medskip\\\textbf{Provenance.} An open information system needs ``hooks'' for information consumers to obtain the trustworthiness of specific information. We believe support for provenance is crucial. 
\medskip\\\textbf{Fully distributed.} A distributed system model is a more robust solution and helps mitigating some of the risk involved in the disaster environment such as partial infrastructure failure. Distributed systems are often characterised along the three axis: consistency, availability, and partitioning. According to the  CAP\footnote{Consistency, Availability and network Partitioning} theorem \cite{cap} any system can only support two out the three desired properties. We believe that in this context, availability and partitioning will be the primary objectives as we don't expect the information stored to be ever consistent.
\medskip\\\textbf{Information availability.} Timely access to the information saves lives. 
The information network needs to guarantee high availability to running on top application. 
\medskip\\\textbf{Support network partitioning.} In a disaster network partitioning is likely to occur.  
The interim report~\cite{inquiry_report} indicates on several occasion system failures that were caused by request overload.  In order to provide a robust and scalable solution the system needs to be distributed and have sufficient resilience to partial failure.  The lack of an ``update'' operator simplifies the re-merging of previously partitioned data sources. 

\section{Related work}\label{related_work}

A report  by the US national Research Council~\cite{national1999reducing} looks into the feasibility of designing a Global Disaster Information Network (GDIN). The report outlines the benefits of GDIN over the current system and identifies potential challenges for implementing such a network. Although inspired by the same goals, this report does not provide any insight on the design or construction of such a network. Nevertheless, it gives a useful list of requirements and justifications for the same. 

The Medical Emergency Disaster Response Network (MEDRN) presented in \cite{medrn2006} is a distributed content-centric network that uses semantic technology to share, disseminate and manage information. The emergency services are able to query for information using a specific query language to obtain the relevant information back from the network without the need for knowing all information providers. All information added to the network becomes instantly available to all users. MERDN uses XML-based content packets for content publication and XML ``interest-profile'' for content subscription. The described proof-of-concept provides a range of useful functionalities such as ``common global view,'' federated search across heterogeneous data sources, real-time update on individual interest. MEDRN, while addressing the same larger objective, focuses mainly on facilitating querying of heterogeneous datasources. Additional issues, such as information context, availability and consistency are not addressed. 

In \cite{trossen2010arguments} Trossen \textit{et al.,} present arguments for information-centric networks. The paper identifies key architectural challenges in implementing such networks and presents a strawman proposal for a future information-centric internetworking  architecture. Our work  is motivated by the same key ideas and challenges presented in the paper, but we take a different approach in the design of GIN.

\section{Global Information Network}\label{GIN}

Motivated by the design goals listed in \ref{design_goals}, we now summarise the key design points behind GIN. For a more detailed description, please refer to~\cite{GIN2011}. 

As depicted in Figure \ref{model}, GIN provides a hypergraph storage abstraction for  applications and services to store and share information.   Each node in GIN represents a specific and unique instance of an ``entity" or a ``value."  Edges and labels respectively define specific relationships between nodes and their types. As mentioned in the introduction, GIN is schema agnostic. It treats the semantics of labels in  nodes and  edges as an end-to-end semantic property.

GIN architecture follows a publish/subscribe paradigm. In contrast to a typical pub/sub system that deals with matching self-contained events against all active subscriptions, GIN operates on a graph. In GIN all publications persist, each publication extends the global graph, while a subscription is created by executing a standing query on the global graph. 

GIN facilitates two basic API functions: \textit{add} and \textit{map}. Application and services using GIN can add to the GIN graph by publishing $n$-tuples. To receive information from GIN, applications utilise standing-graph queries, which essentially maintain a mapping between the GIN's global graph and the applications' local graph. Any information added locally is reflected in GIN and any changes to the global graph made by others are reflected back into application's state if it falls within the defined mapping.  %

\subsection{Disaster Management with GIN}

In this section we present a scenario following Figure \ref{gin_based_model}, on how the disaster management communication can be done using GIN. 

We assume that GIN already contains information from  
the deployed rainfall and river level gauges. Using the GIN API, the real-time monitoring system will retrieve the required information by subscribing to all the gauges within the relevant catchment area. It will also periodically produce an analysis and publish the result back into GIN. %

Analysing and visualising applications or services can be easily developed by subscribing to the information published by the real-time monitoring system. 
Based on this information, the flood engineers will devise the corresponding gate strategy,  and publish it back to GIN. 
In turn, the councils' systems will be able to subscribe to the ``situation reports'' published by the flood engineers and make the relevant decision. These decisions will be published back into the GIN for historical record and/or future reference. Since everyone will share the space information network, the council will also be able to subscribe to the warning and forecast from the BoM and add annotations regarding the impact on the local community.

Local web servers, social media website can use a subscription to display the most recent information or  warnings added by the councils. Public comments from  websites or social networking will be also captured within GIN. As a result the community responses will appear in context, linked to other information. For example,  comments from local residents posted on social networks can be integrated into the incoming information flow by the interested agencies such as emergency response units that can use it to identify hotspots that require urgent attention.

\subsection{Observation}
There are many disaster management stakeholders at different levels that produce, collect and process information in an autonomous and largely distributed manner. While advances in technology lead to a constant increase in new data generated in varied formats, the underlying infrastructure provides little support other than transferring data.  Although the described scenario can be implemented today by glueing or meshing various services together, we would like to note the ease with which it can be realised on the GIN and more importantly, how seamlessly  it can be extended without any modification to the base service. 

The GIN abstraction facilitates a more efficient interaction and coordination among autonomous agents through a shared information space. A system running on top of GIN will be comprised of a collection of agents: ``simple'' that publish data into the GIN and/or more complex (`smart') that interact with the external environment, subscribe to relevant information within GIN, make calculation and publish the resulting information back into GIN for others to use. No information is removed from GIN. This becomes extremely useful, in the recovery and preparation stages, when agencies will want to learn from past events and  need to query for previously published information.

The information-centric approach allows tapping into new and previously untouched resources.  GIN's underlying hypergraph abstraction facilitates a way to instill information on information. In aforementioned disaster scenario, data is added to GIN from multiple sources with  reference to its origin and other related information.  %
For example, while publishing their ``situation reports'' FOC's flood engineers can link them to previously published gauges data and BoM rainfall predictions, providing the ultimate decision-makers with a more educated view on the data they get \textit{, e.g.,} where it comes from and who produced it?

GIN opens new avenues in disaster management,  such as the utilisation of ``crowdsourcing'' techniques to get a more complete picture of the evolving events on the ground. 

We envision that GIN will be able to provide more enhanced functionality to currently available systems and open ways to develop new type of applications that take advantage the global information space, for example, various CROP-types systems. 

\section{Current Progress}\label{challenge_and_progress}

We are working on a large-scale tuple store based on Distributed Hash Tables (DHT) as a key building block for GIN. The content of the GIN can be fully described by a list of 7-tuples, using UUIDs\footnote{Universally Unique IDentifier} as vertices and edge identifiers. Each tuple contains the two vertices, the connecting edge, the context vertex (to define hypergraphs), a timestamp, the optional signer (an entity itself) and respective signature over the tuple.

The tuple store supports \textit{add} and \textit{map} functions.
The {\em add} function  is a simple insertion  of a sequence of tuples to the tuple store. 
The {\em map} API is slightly more complex as it creates a stateful context. It is essentially a standing  graph query on a tuple store that will result in an incoming stream of tuples. We represent any graph query  as a set of tuple templates on all available tuples in the GIN. Joins are performed at the end device. 
To facilitate standing graph queries and efficient real-time notification support in our implementation, we have adapted the Rete algorithm~\cite{rete}. Rete generates a dataflow network from a given set of rules or standing  queries, identifying all common sub goals. The algorithm's chain of successively triggered right and left activations minimises the computational load and locale  for each publication given a global set of subscriptions.  

Specifically, our tuple store performs the role of the $\alpha$  network in the Rete algorithm as it provides the same template based subscription service. %
To this end, we extended the Kademlia DHT \cite{kademlia} with a \textit{multi\_get} operation. The \textit{ multi\_get} operation accepts bit-vector patterns  with wildcards to retrieve all matching tuples. For more information refer to~\cite{yan:sigcomm11:poster}. Joining across  tuple streams, as provided by Rete's $\beta$ network can be  realised in different ways depending on the architecture's objectives. Our prototype currently implements this on  the end host in a purely pull driven manner. 

This simple filter mechanism clearly has performance implications concerning the number of tuples which will get dropped in subsequent joins as well as the lack of flow control on the receiving end. There are many solutions in the literature to address these problems but our initial design objective was to keep it as simple as possible.

\section{Conclusions and future work}

In this paper we have presented a case for the adoption of an information-centric architecture for a global disaster management system.  

We have used the 2010/2011 Queensland floods case study to describe challenges in providing every participant with relevant and actionable information. We used various examples to motivate a more flexible information dissemination framework which is designed to quickly add and use new information while facilitating information quality assurance. 

We then listed a number of key requirements for an information centric model and briefly described our proposed architecture and and gave a brief account of our current progress.

Our future work will involve evaluating the above architecture and various design decisions through deployment of actual services on real networks, such as  PlanetLab \cite{planetlab} or GENI \cite{GIN2011}. 

\small
\bibliographystyle{elsarticle-num}
\bibliography{references}

\begin{thebibliography}{10}
\expandafter\ifx\csname url\endcsname\relax
  \def\url#1{\texttt{#1}}\fi
\expandafter\ifx\csname urlprefix\endcsname\relax\def\urlprefix{URL }\fi
\expandafter\ifx\csname href\endcsname\relax
  \def\href#1#2{#2} \def\path#1{#1}\fi

\bibitem{GIN2011}
M.~{Ott}, Y.~{Shvartzshnaider}, \href{http://arxiv.org/abs/1104.2134v1}{{{A
  Case for a Global Information Network}}}, ArXiv e-prints\href
  {http://arxiv.org/abs/1104.2134} {\path{arXiv:1104.2134}}.
\newline\urlprefix\url{http://arxiv.org/abs/1104.2134v1}

\bibitem{lod_cloud}
Strategies for building semantic web applications,
  \url{http://notes.3kbo.com/linked-data} (October 2011).

\bibitem{national1999reducing}
N.~R. C.~B. on~Natural~Disasters, Reducing Disaster Losses Through Better
  Information, Natl Academy Pr, 1999.

\bibitem{qdm_website}
Queensland's disaster management website, \url{http://www.disaster.qld.gov.au/}
  (August 2011).

\bibitem{inquiry_report}
Queensland floods commission of inquiry,
  \url{http://www.floodcommission.qld.gov.au} (August 2011).

\bibitem{brisbane_report}
{Independent Review of Brisbane City Council's Response, 9-22 January 2011.},
  \url{http://www.brisbane.qld.gov.au/} (August 2011).

\bibitem{cap}
E.~Brewer, Towards robust distributed systems, in: Proceedings of the Annual
  ACM Symposium on Principles of Distributed Computing, Vol.~19, 2000, pp.
  7--10.

\bibitem{medrn2006}
J.~Gomezjurado, D.~Reininger, Medrn-a mutual aid information network for
  emergency response, in: Military Communications Conference, 2006. MILCOM
  2006. IEEE, IEEE, 2006, pp. 1--7.

\bibitem{trossen2010arguments}
D.~Trossen, M.~Sarela, K.~Sollins, Arguments for an information-centric
  internetworking architecture, ACM SIGCOMM Computer Communication Review
  40~(2) (2010) 26--33.

\bibitem{rete}
R.~B. Doorenbos, Production matching for large learning systems, Ph.D. thesis,
  Citeseer (1995).

\bibitem{kademlia}
P.~Maymounkov, D.~Mazieres, Kademlia: A peer-to-peer information system based
  on the xor metric, in: Proc. of {IPTPS02,} Cambridge, {USA}, Vol.~1, 2002,
  pp. 2--2.

\bibitem{yan:sigcomm11:poster}
Y.~Shvartzshnaider, M.~Ott, Towards a fully distributed n-tuple store, in:
  Proceedings of the ACM SIGCOMM 2011 conference on SIGCOMM, ACM, 2011, pp.
  388--389.

\bibitem{planetlab}
B.~Chun, D.~Culler, T.~Roscoe, A.~Bavier, L.~Peterson, M.~Wawrzoniak,
  M.~Bowman, Planetlab: an overlay testbed for broad-coverage services, ACM
  SIGCOMM Computer Communication Review 33~(3) (2003) 3--12.

\end{thebibliography}

\end{document}